\documentclass[%
 reprint,
superscriptaddress,
 amsmath,amssymb,
 aps,prc
]{revtex4-2}

\usepackage{graphicx}
\usepackage{dcolumn}
\usepackage{bm}
\usepackage[output-decimal-marker={.},exponent-product=\cdot]{siunitx}
\usepackage{hyperref}
\usepackage[mathlines]{lineno}
\usepackage{makecell}
\usepackage{color}
\usepackage[english]{babel}
\usepackage[T1]{fontenc}
\usepackage{multirow}
\usepackage{float}

\begin{document}

\preprint{APS/123-QED}

\title{Probing the N=104 midshell region for the r process via precision mass spectrometry of neutron-rich rare-earth isotopes with the JYFLTRAP double Penning trap}

\author{A.~Jaries}
\email{arthur.a.jaries@jyu.fi}
\affiliation{University of Jyvaskyla, Department of Physics, Accelerator laboratory, P.O. Box 35(YFL) FI-40014 University of Jyvaskyla, Finland}
\affiliation{Helsinki Institute of Physics, FI-00014, Helsinki, Finland}
\author{S.~Nikas}
\email{stylianos.s.nikas@jyu.fi}
\affiliation{University of Jyvaskyla, Department of Physics, Accelerator laboratory, P.O. Box 35(YFL) FI-40014 University of Jyvaskyla, Finland}
\author{A.~Kankainen}
\email{anu.kankainen@jyu.fi}
\affiliation{University of Jyvaskyla, Department of Physics, Accelerator laboratory, P.O. Box 35(YFL) FI-40014 University of Jyvaskyla, Finland}
\author{T.~Eronen}
\affiliation{University of Jyvaskyla, Department of Physics, Accelerator laboratory, P.O. Box 35(YFL) FI-40014 University of Jyvaskyla, Finland}
\author{O.~Beliuskina}
\affiliation{University of Jyvaskyla, Department of Physics, Accelerator laboratory, P.O. Box 35(YFL) FI-40014 University of Jyvaskyla, Finland}
\author{T.~Dickel}
\affiliation{GSI Helmholtzzentrum f\"ur Schwerionenforschung GmbH, 64291 Darmstadt, Germany}
\affiliation{II. Physikalisches Institut, Justus Liebig Universit\"at Gie{\ss}en, 35392 Gie{\ss}en, Germany}
\author{M.~Flayol}
\affiliation{Universit\'e de Bordeaux, CNRS/IN2P3, LP2I Bordeaux, UMR 5797, F-33170 Gradignan, France}
\author{Z.~Ge}
\affiliation{University of Jyvaskyla, Department of Physics, Accelerator laboratory, P.O. Box 35(YFL) FI-40014 University of Jyvaskyla, Finland}
\affiliation{GSI Helmholtzzentrum f\"ur Schwerionenforschung GmbH, 64291 Darmstadt, Germany}
\author{M.~Hukkanen}
\affiliation{University of Jyvaskyla, Department of Physics, Accelerator laboratory, P.O. Box 35(YFL) FI-40014 University of Jyvaskyla, Finland}
\affiliation{Universit\'e de Bordeaux, CNRS/IN2P3, LP2I Bordeaux, UMR 5797, F-33170 Gradignan, France}
\author{M.~Mougeot}
\affiliation{University of Jyvaskyla, Department of Physics, Accelerator laboratory, P.O. Box 35(YFL) FI-40014 University of Jyvaskyla, Finland}
\author{I.~Pohjalainen}
\affiliation{University of Jyvaskyla, Department of Physics, Accelerator laboratory, P.O. Box 35(YFL) FI-40014 University of Jyvaskyla, Finland}
\author{A.~Raggio}
\affiliation{University of Jyvaskyla, Department of Physics, Accelerator laboratory, P.O. Box 35(YFL) FI-40014 University of Jyvaskyla, Finland}
\author{M.~Reponen}
\affiliation{University of Jyvaskyla, Department of Physics, Accelerator laboratory, P.O. Box 35(YFL) FI-40014 University of Jyvaskyla, Finland}
\author{J.~Ruotsalainen}
\affiliation{University of Jyvaskyla, Department of Physics, Accelerator laboratory, P.O. Box 35(YFL) FI-40014 University of Jyvaskyla, Finland}
\author{M.~Stryjczyk}
\affiliation{University of Jyvaskyla, Department of Physics, Accelerator laboratory, P.O. Box 35(YFL) FI-40014 University of Jyvaskyla, Finland}
\author{V.~Virtanen}
\affiliation{University of Jyvaskyla, Department of Physics, Accelerator laboratory, P.O. Box 35(YFL) FI-40014 University of Jyvaskyla, Finland}

\begin{abstract}
 We have performed high-precision mass measurements of neutron-rich rare-earth Tb, Dy and Ho isotopes using the Phase-Imaging Ion-Cyclotron-Resonance technique at the JYFLTRAP double Penning trap. We report on the first experimentally determined mass values for $^{169}$Tb, $^{170}$Dy and  $^{171}$Dy, as well as the first high-precision mass measurements of $^{169}$Dy and $^{169\text{-}171}$Ho. For $^{170}$Ho, the two long-lived ground and isomeric states were resolved and their mass measured, yielding an isomer excitation energy of $E_\text{exc}=150.8(54)$~keV. In addition, we have performed independent crosschecks of previous Penning-trap values obtained for $^{167\text{,} 168}$Tb and $^{167\text{,} 168}$Dy. We have extended the systematics of two-neutron separation energies to the neutron midshell at $N=104$ in all of the studied isotopic chains. Our updated and new mass measurements provide better mass-related constraints for the neutron-capture reaction rates relevant to the astrophysical rapid neutron capture (r) process. The r-process abundances calculated with the new mass values seem to produce a steeper minimum at A=170 and differ by around 15-30\% from the abundances computed with the Atomic Mass Evaluation 2020 values.
\end{abstract}

\maketitle
\section{\label{sec:intro}Introduction}
The properties of neutron-rich rare-earth isotopes are important both for studying how the nuclear structure evolves further from stability, and for better understanding of the astrophysical rapid neutron capture process (the r process) \cite{RevModPhys.93.015002}. These isotopes are located between the major closed proton and neutron shells at $Z=50$ and 82, and $N=82$ and 126. A strong onset of deformation has been observed to take place at $N\approx88-90$ \cite{brix1952nuclear,mottelson1955classification} and been verified to continue toward larger neutron numbers via $\gamma$-ray spectroscopy (see e.g. Refs. \cite{Patel2014,Wu2017}). 
An interesting question is related to what happens at the midshell at N=104. Strong structural changes at that neutron number have been experimentally observed in neutron-deficient isotopes in the lead region, see Ref. \cite{Garrett2022} and references therein. As suggested in Ref.~\cite{surman1997source}, a presence of such effects could affect neutron separation energies and calculated r-process abundances.
In the $N\approx66$ midshell region, recent Penning-trap measurements have indicated somewhat lower two-neutron separation energies $S_{2n}$ than expected from a linearly decreasing trend for the Zr ($Z=40$) \cite{Hukkanen2024} and Sr ($Z=38$) \cite{Mukul2021} isotopic chains at the neutron midshell $N=66$. No such data exist yet for the $Z\approx64,N=104$ midshell region.

The r process produces around half of the elemental abundances heavier than iron \cite{RevModPhys.29.547}. The mechanisms involved in the rare-earth abundance peak formation are still not fully understood. It has been proposed to be related to a nearby subshell closure or strong nuclear deformation in the region \cite{surman1997source,mumpower2012formation}, double asymmetric fission fragment distribution in the fission recycling from heavier nuclei \cite{goriely2013new}, or a combination of these scenarios. Nevertheless, the rare-earth abundance peak is very sensitive to the properties of neutron-rich nuclei in the $N=104$ neutron midshell region \cite{mumpower2012formation,mumpower2012influence,mumpower2015impact,hao2023sensitivity}, in particular to nuclear masses. Reducing mass-related uncertainties gives better constraints on the calculated astrophysical reaction rates and hence, on the calculated r-process abundance pattern. 

Previously, isotopic chains from neodymium ($Z=60$) to dysprosium ($Z=66$) have been  explored via mass measurements at JYFLTRAP and at Canadian Penning Trap (CPT) \cite{vilen2020exploring,vilen2018precision,orford2018precision,orford2019phase}. However, the measurements did not yet reach the N=104 midshell 
and up to now the determined $S_{2n}$ followed a smooth linear trend. Thus, no strong evidence of structural changes or subshell closures was discovered. To confidently extend the available experimental data towards measurements of more exotic isotopes in these isotopic chains, consistency checks with existing Penning trap measurements are needed to benchmark our experimental setup.

No high-precision mass measurements of radioactive neutron-rich holmium ($Z=67$) isotopes, the next heavier isotopic chain, have been performed despite their astrophysical interest. $^{170}$Ho$^\text{m}$ is an astromer candidate \cite{misch2020astromers}, a nuclear isomer that is predicted to have a significant impact on the astrophysical neutron-capture reaction rates, and $^{171}$Ho, the most exotic holmium with a known mass, is reported with a large uncertainty on its mass excess (ME$_\text{lit.}=\num{-54518(600)}~$keV \cite{AME2020}). Thus, there is a need to improve the $^{171}$Ho mass value via high-precision Penning trap mass spectrometry \cite{Eronen2016,Dilling2018}. 

In this work, we report on the first high-precision mass measurements extending to the $N=104$ neutron midshell in the rare-earth region, namely $^{167\text{-}169}$Tb, $^{167\text{-}171}$Dy, and $^{169\text{-}171}$Ho determined with the Phase-Imaging Ion-Cyclotron-Resonance (PI-ICR) method \cite{eliseev2013phase,eliseev2014phase,nesterenko2018phase,nesterenko2021study} at the JYFLTRAP double Penning trap \cite{kolhinen2004jyfltrap,eronen2014jyfltrap}. We discuss the impact of these measurements on the astrophysical neutron capture reaction rates, r-process abundances and nuclear structure in the $N=104$ region.

\section{Experimental methods}
The mass measurements of neutron-rich rare-earth isotopes around $N=104$ were performed at the Ion Guide Isotope Separator On-Line (IGISOL) facility \cite{moore2013towards,gao2020fission}. The ions of interest were produced via proton-induced fission on a thin natural uranium target (15 mg/cm$^2$ thickness) using a 25-MeV proton beam provided by the K-130 cyclotron with an intensity ranging from 1 to 22 $\mu$A. The fission products were first thermalized in a gas cell filled with 300 mbar of helium buffer gas, from which they were then guided out by the gas flow and extracted through a Sextupole Ion Guide (SPIG) \cite{karvonen2008sextupole}, mostly as singly-charged ions. 

The ions were accelerated to 30$q$~keV and separated based on their mass-over-charge ratio $m/q$ using a dipole magnet. The selected ions were continuously injected into the Radiofrequency Quadrupole Cooler Buncher (RFQCB) \cite{nieminen2001beam}, where they were cooled and then extracted out as ion bunches. After the RFQCB, the ions were finally sent towards the JYFLTRAP double Penning trap mass spectrometer \cite{kolhinen2004jyfltrap,eronen2014jyfltrap}. 

In the JYFLTRAP, the captured ions were electromagnetically confined, radially by a strong homogeneous magnetic field $B$ in the beam axis direction using a 7~T superconducting magnet, and axially by a 100~V deep quadrupole potential. In the first trap (purification trap), filled with helium gas, the ions are cooled, centered, and the ion of interest selected by using the buffer gas cooling technique \cite{savard1991new}. The latter allows cleaning the ion bunch from most of the isobaric contamination with a resolving power of typically $R>5\times 10^4$. The purified ion bunch is then transferred to the second trap (precision trap) where the free space cyclotron frequency $\nu_c$ of the species of interest, defined as,
\begin{equation}
    \nu_c=\frac{1}{2\pi}\frac{q}{m}B
\end{equation}
was determined using the Phase-Imaging Ion Cyclotron Resonance (PI-ICR) technique \cite{eliseev2013phase,eliseev2014phase,nesterenko2018phase,nesterenko2021study}. 

\begin{figure}[h!t!b]
\includegraphics[width = 0.5\textwidth]{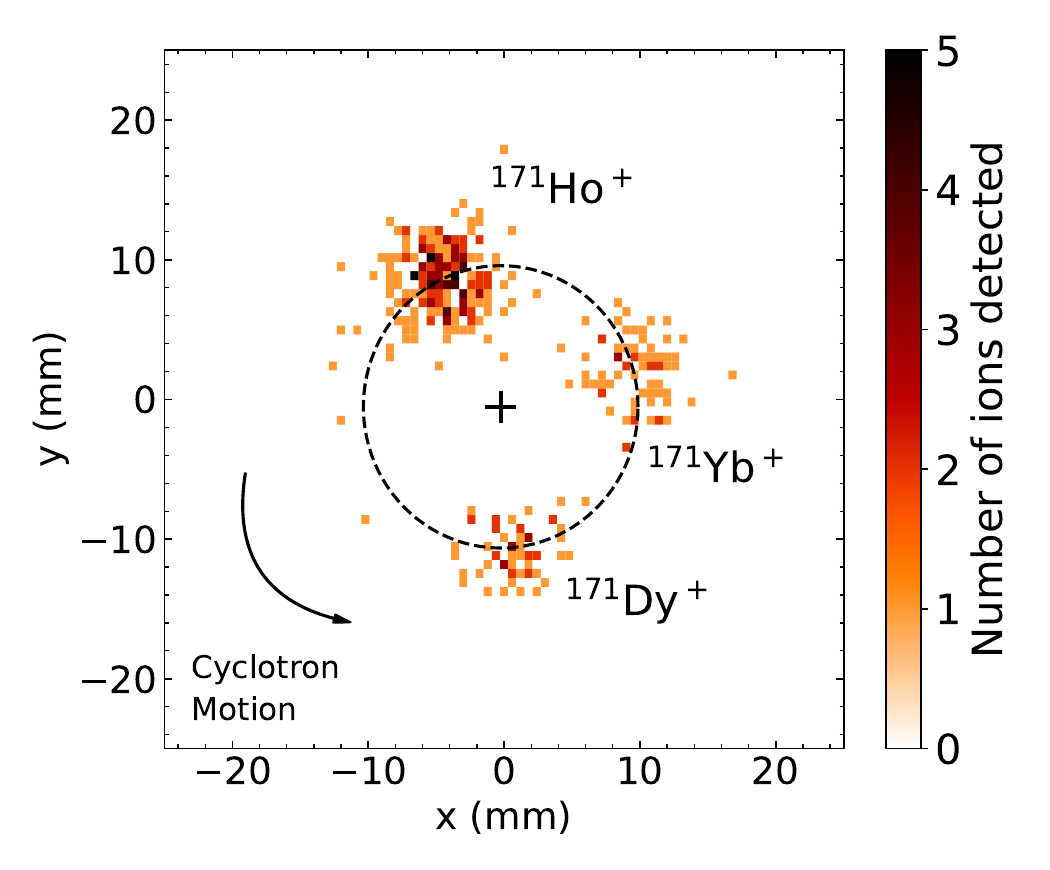}
\caption{Cyclotron motion projection of $^{171}$Dy$^+$, $^{171}$Ho$^+$ and $^{171}$Yb$^+$ ions on the 2D position-sensitive detector with the PI-ICR technique. The three cyclotron phase spots are separated using a 258~ms phase accumulation time. The center position is marked with the + symbol. The approximated motion orbit is indicated with the dashed circle.  \label{fig:piicr_dy} }
\end{figure}

The cyclotron frequency $\nu_c$ was derived as follows,
\begin{equation}
    \nu_c=\frac{\phi_c+2\pi (n_\text{-}+n_\text{+})}{2\pi t_\text{acc}}
\end{equation}
where $\phi_c$ is the phase difference between the respective phases of the two coupled radial eigenmotions of the trapped ions, namely magnetron ($\phi_\text{-}$) and reduced cyclotron ($\phi_\text{+}$). The angles were detected on a position sensitive detector (a 2D-MCP) upon extraction after a given accumulation time $t_\text{acc}$. An example of a PI-ICR figure obtained in this work is presented in Fig.~\ref{fig:piicr_dy}. By choosing an accumulation time to be a multiple integer of $\nu_c$, $\phi_c$ is kept small in order to minimize the image distortion shift \cite{nesterenko2021study}. In order to avoid the $\phi_+$ motion projections of possible isobaric, isomeric, and molecular contaminants overlapping with the ion of interest, $t_\text{acc}$ is also chosen so all the ion species in presence are clearly separated on the position sensitive detector. Different accumulation times are used prior to the mass measurement for an unambiguous identification of the species, especially in cases where the uncertainties from the literature are large. The integers $n_\text{-}$ and $n_\text{+}$ correspond to the number of turns the ions perform in each in-trap radial eigenmotion during $t_\text{acc}$. To reliably determine the phase difference, the center position of the precision trap was regularly determined over the mass measurement by extracting the trapped ions without excitation of their eigenmotions. 

To obtain a precise measurement of the magnetic field $B$ over time, a similar measurement using a reference ion with a precisely known mass was performed alternately. The reference ions were $^{133}$Cs$^+$, $^{170}$Ho$^+$ or stables $^{170, 171}$Yb$^+$. They were either, produced from the IGISOL off-line surface ion source \cite{vilen2020new}, directly coming from the proton-induced fission reaction, or impurities ionized in the surroundings of the ion guide, respectively. The mass value for the  singly-charged ion of interest $m_\text{ion}$, was finally calculated from the ratio of the determined cyclotron frequencies $r=\nu_{c\text{,ref}}/\nu_{c\text{,ion}}$ according to
\begin{equation}
    m_\text{ion}=(m_\text{ref}-m_e)r+m_e,
\end{equation}
where $m_\text{ref}$ is the atomic mass of the chosen reference and $m_e$ the mass of a free electron. The excitation energy $E_\text{exc}$ of an isomeric state measured with respect to its ground state as a reference was extracted as follows:
\begin{equation}
\label{eq:Q}
    E_\text{exc}=(m_\text{isomer}-m_\text{gs})c^2=(m_\text{ref}-m_e)(r-1)c^2  \text{,}
\end{equation}
where $c$ is the speed of light in vacuum.  The analysis accounted for the error related to the ion motion projection distortion, and for the magnetron phase advancement correction in case multiple ion species were still present in the precision trap \cite{nesterenko2021study}. The contribution from electron binding energies was neglected as of the order of a few eV. Due to the low statistics gathered on the measured cases, no count-rate class analysis \cite{nesterenko2021study,kellerbauer2003direct} was performed. A gate at one detected ion per bunch was set instead to limit the influence of possible ion-ion interactions. Systematic uncertainties related to the magnetic field fluctuations over the time of acquisition were taken into account following $\delta B/B = 2.01(25) \times 10^{-12}$ min$^{-1}$. A mass-dependent uncertainty of $\delta_m r/r = -2.35(81) \times 10^{-10} / \textnormal{u} \times (m_\text{ref} - m_\text{ion})$ and a residual of $\delta_\text{res}r/r=9\times 10^{-9}$ \cite{nesterenko2021study} were added when the reference used for the measurement was not isobaric with respect to the ion of interest.

\begin{table*} [h!t!b]
\caption{\label{tab:table3}List of the nuclides studied in this work with their respective half-lives $T_{1/2}$ and spin parities $J^\pi$ taken from NUBASE20 \cite{NUBASE2020}. The used reference nuclide (Ref.) and accumulation time $t_{acc}$ used to perform the PI-ICR measurement, as well as the obtained frequency ratios $r=\nu_\text{c,ref}/\nu_\text{c,ion}$ are tabulated. The mass excesses from this work, $\text{ME}_\text{JYFLTRAP}$, are reported together with the literature values $\text{ME}_\text{lit.}$, taken from the AME20 \cite{AME2020}, and their difference $\text{ME}_\text{JYFLTRAP}-\text{ME}_\text{lit.}$ (Diff.) is calculated. The $\#$ symbol refers to extrapolated values from systematics.}
\begin{ruledtabular}
\begin{tabular}{llllllllll}
  Nuclide & $T_{1/2}$  & $J^{\pi}$ & Ref. & $t_{acc}$ (ms) & $r=\nu_\text{c,ref}/\nu_\text{c,ion}$ & $\text{ME}_\text{JYFLTRAP}$ (\text{keV}) & $\text{ME}_\text{lit.}$ (\text{keV})& \text{Diff.} (\text{keV})\\ \hline
\: $^{167}\text{Tb}$  & $18.9(16)$ s     &  (3/2$^+$) & $^{133}\text{Cs}$ & 448 & \num{1.256081995(38)} & \num{-55885.2(47)} & \num{-55883.1(19)} & $-2(5)$\\
\: $^{168}\text{Tb}$  & $9.4(4)$ s      &    (4$^-$) & $^{133}\text{Cs}$ & 520 & \num{1.263631522(37)} & \num{-52746.8(46)} & \num{-52781(4)} & 34(6)\\
\: $^{169}\text{Tb}$     & $5.13(32)$ s     &   3/2$^+\#$    & $^{133}\text{Cs}$ & 432 & \num{1.271174910(77)} & \num{-50368.5(96)} & \num{-50480(300)}\# & 112(300)\#\\
\: $^{167}\text{Dy}$     & $6.20(8)$ min & (1/2$^-$) & $^{133}\text{Cs}$ & 448 & \num{1.256049417(35)} & \num{-59918.4(44)} & \num{-59911(4)} & $-7(6)$\\
\: $^{168}\text{Dy}$     & 8.7(3) min    & 0$^+$     & $^{133}\text{Cs}$ & 520 & \num{1.263585172(32)} & \num{-58485.0(40)} & \num{-58560(140)}\footnotemark[1] & 75(140)\\
\: $^{169}\text{Dy}$     & 39(8) s       & (5/2)$^-$ & $^{133}\text{Cs}$ & 432 & \num{1.271133274(46)} & \num{-55523.1(57)}& \num{-55600(300)} & 77(300)\\
\: $^{170}\text{Dy}$     &    54.9(80) s    & 0$^+$     & $^{170}\text{Yb}$ & 368 & \num{1.000044533(41)} & \num{-53714.6(63)} & \num{-53710(300)}\# & $-5(300)\#$\\
\: $^{171}\text{Dy}$     & $4.1(4)$ s     &  7/2$^-\#$   & $^{171}\text{Yb}$ & 258 & \num{1.000057356(73)} & \num{-50174(12)} & \num{-50010(200)}\# & $-164(200)\#$\\
\: $^{169}\text{Ho}$     & 4.72(10) min  & 7/2$^-$   & $^{133}\text{Cs}$ & 642 & \num{1.271106638(38)} & \num{-58820.7(47)} & \num{-58796(20)} & $-25(21)$\\
\: $^{170}\text{Ho}$     & 2.76(5) min   & (6$^+$)   & $^{170}\text{Yb}$ & 368 & \num{1.000028429(49)} & \num{-56263.8(78)} & \num{-56240(50)} & $-24(51)$\\
\: $^{170}\text{Ho}^\text{m}$ & 43(2) s       & (1$^+$)   & $^{170}\text{Ho}$ & 368 & \num{1.000000953(34)}\footnotemark[2] & \num{-56113.0(95)} & \num{-56140(60)} & 27(61)\\
\: $^{171}\text{Ho}$     & 53(2) s       & 7/2$^-\#$ & $^{171}\text{Yb}$ & 258 & \num{1.000029506(46)} & \num{-54608.8(74)} &\num{-54520(600)} & $-89(600)$\\
\end{tabular}
\end{ruledtabular}
\footnotetext[1]{Also \num{-58486.0(40)} keV in Ref. \cite{orford2019phase}}
\footnotetext[2]{This gives an excitation energy of $E_\text{exc}=150.8(54)$~keV}
\end{table*}

\section{Results and discussion}
A summary of the experimental values reported in this work can be found in Table~\ref{tab:table3}. Detailed discussion is presented in the following subsections.

\subsection{$^{167\text{-}169}$Tb \label{terbium}}
The neutron-rich terbium ($Z=65$) isotopes were all measured against $^{133}$Cs$^+$ ions. While the mass excess of $^{167}$Tb we report in this work, ${\text{ME}= -55885.2(47)}$~keV, agrees with the Atomic Mass Evaluation 2020 (AME20) \cite{AME2020}, the mass measurement of $^{168}$Tb shows a 34~keV shift ($\sim6\sigma$). A reason for this discrepancy can be explained by the presence of $^{168}$Dy contamination in the TOF-ICR resonances of Ref.~\cite{vilen2020exploring}. 

The $^{168}$Tb mass measurement was performed using the TOF-ICR technique \cite{graff1980direct,konig1995quadrupole} enhanced with the Ramsey method of time-separated oscillatory fields \cite{kretzschmar2007ramsey,george2007ramsey}. A 25-750-25~ms (On-Off-On) Ramsey excitation pattern was chosen to separate the converted $^{168}$Tb ions from the neighbouring isobar $^{168}$Dy, assuming the mass excess of the latter to be $\text{ME}_\text{lit}=-58560(140)$~keV from the AME2016 \cite{wang2017ame2016}. In this work, we performed another Penning-trap mass measurement of $^{168}$Dy indicating a mass excess of 75~keV higher than reported in the AME \cite{wang2017ame2016,AME2020}. Considering this new mass-excess value, the 25-750-25~ms (On-Off-On) Ramsey excitation pattern presents an overlap of the side minimums of the two species, which prevents the contamination from being clearly observed in the data analysis. As the PI-ICR technique used in this work offers a greater resolving power, $^{168}$Tb was unambiguously separated from $^{168}$Dy and its mass excess measured to be $\text{ME}=-58746.8(46)$~keV.

\begin{figure}[h!t!b]
\includegraphics[width = 0.5\textwidth]{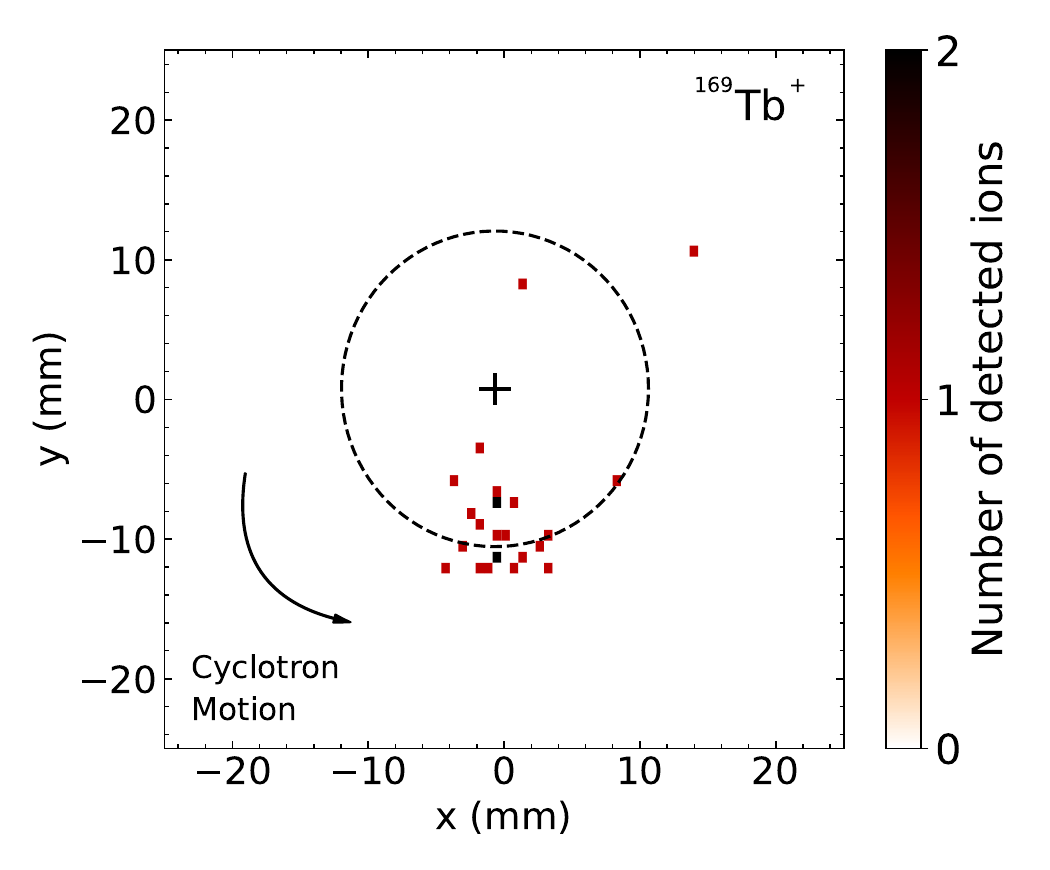}
\caption{Cyclotron motion projection of $^{169}$Tb$^+$ ions on the 2D position-sensitive detector with the PI-ICR technique. The ion motion has been accumulated for 432 ms. The center position is marked with the + symbol. The approximated motion orbit is indicated with the dashed circle.\label{fig:piicr_tb} }
\end{figure}

The first direct mass measurement of $^{169}$Tb was performed in this work. The measured mass-excess value is ${\text{ME} = -50368.5(96)}$~keV and the measurement, performed with a $432$~ms accumulation time, see Fig.~\ref{fig:piicr_tb}.

\subsection{$^{167\text{-}171}$Dy \label{dysprosium}}
In this work we explored the neutron-rich dysprosium isotopic chain, extending the experimentally determined masses up to $^{171}$Dy ($N=105$). The extracted mass-excess value of  $^{167}$Dy, ${\text{ME} = -59918.4(44)}$~keV, lies $7(6)$~keV lower ($1.2\sigma$) than reported from the previous Penning-trap measurement performed at CPT \cite{orford2019phase} and included in the AME20 \cite{AME2020} evaluation. 

For $^{168}$Dy, the literature value, ${\text{ME}_\text{lit.} = -58560(140)}$~keV \cite{AME2020}, is based on a heavy-ion transfer-reaction experiment \cite{lu1998study}. However, a more recent and precise CPT measurement \cite{orford2019phase} reports a mass-excess value of ${\text{ME}_\text{CPT} = -58486.0(40)}$~keV \cite{orford2019phase} 
with which our value, ${\text{ME} = -58485.0(40)}$~keV, is in perfect agreement.

Our mass value of $^{169}$Dy, ${\text{ME} = -55523.1(57)}$~keV, agrees with the literature value ${\text{ME}_\text{lit.} = -55600(300)}$~keV \cite{AME2020} evaluated from a multinucleon transfer reaction study \cite{chasteler1990decay}. The precision of the mass-excess value was improved by a factor of 53 in this work.

This work reports the first direct mass measurements of the doubly midshell nucleus $^{170}$Dy, and  $^{171}$Dy (${\text{ME} = -53714.6(63)}$~keV and ${\text{ME} = -50174(12)}$~keV, respectively). Both isotopes were measured with a 258~ms and 368~ms phase accumulation times, and with respect to stable isobars $^{170}$Yb (${\text{ME}_\text{lit.} = -60763.929(10)}$~keV) and $^{171}$Yb  (${\text{ME}_\text{lit.} = -59396.818(12)}$~keV), respectively. The obtained PI-ICR measurement of $^{171}$Dy is shown in Fig.~\ref{fig:piicr_dy}.

\subsection{$^{169\text{-}171}$Ho \label{holmium}}
The measured mass of $^{169}$Ho, ${\text{ME} = -58820.7(47)}$~keV, is 25(21)~keV lower ($\sim  1.2\sigma$) and four times more precise than the literature value ${\text{ME}_\text{lit.} = -58796(20)}$~keV \cite{AME2020}. The AME20 value is based on the multinucleon transfer-reaction study performed at the OASIS facility \cite{chasteler1990decay}.

\begin{figure}[h!t!b]
\includegraphics[width = 0.5\textwidth]{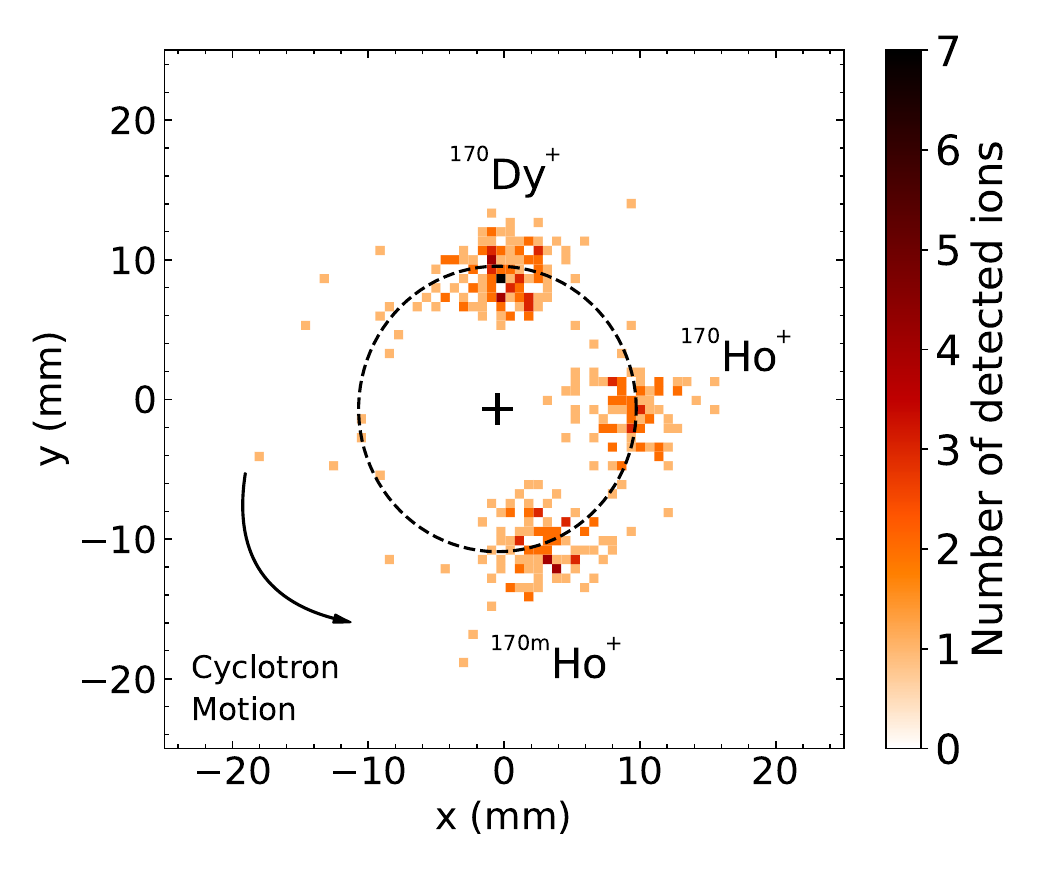}
\caption{Cyclotron motion projection of $^{170}$Dy$^+$, $^{170}$Ho$^+$ and $^{170\text{m}}$Ho$^+$ ions on the 2D position-sensitive detector with the PI-ICR technique. The three cyclotron phase spots are separated using a 368 ms phase accumulation time. The center position is marked with the + symbol. The approximated motion orbit is indicated with the dashed circle. \label{fig:piicr_ho} }
\end{figure}

In this work, the $^{170}$Ho ground state mass-excess value was determined using $^{170}$Yb as a reference while the long-lived isomeric state, $^{170}$Ho$^\text{m}$, was measured with respect to the ground state. Using Eq.~\ref{eq:Q}, the excitation energy of the isomer was determined to be $E_\text{exc}=150.8(54)$~keV. This is 51(80)
~keV above the literature value ($E_\text{exc, lit.}=100(80)$~keV \cite{NUBASE2020}) and 15 times more precise. 
Both the ground and isomeric state mass-excess values of $^{170}$Ho agree with the literature  values \cite{NUBASE2020}, which are based on a $\beta$-decay experiment \cite{tuurnala1978energies}. The precision of the mass-excess values for $^{170}$Ho and  $^{170}$Ho$^\text{m}$ were improved in this work by a factor of six and eight, respectively. The mass measurements of both states were performed with a 368~ms accumulation time and is presented in Fig.~\ref{fig:piicr_ho}.

$^{171}$Ho constitutes the heaviest $^\text{nat}$U$(p,f)$ fission fragment measured to date with the JYFLTRAP Penning trap at IGISOL \cite{moore2013towards,gao2020fission}. Its mass excess, ${\text{ME} = -54608.8(74)}$~keV, was measured against the stable isobar $^{171}$Yb. The new mass-excess value is 89~keV lower and 81 times more precise than the literature value \cite{AME2020}, ${\text{ME}_\text{lit.} = -54520(600)}$~keV, which is based on the previously mentioned multinucleon transfer experiment \cite{chasteler1990decay}.

\subsection{Two-neutron separation energies}
\begin{figure}[h!t!b]
\includegraphics[width = 0.45\textwidth]{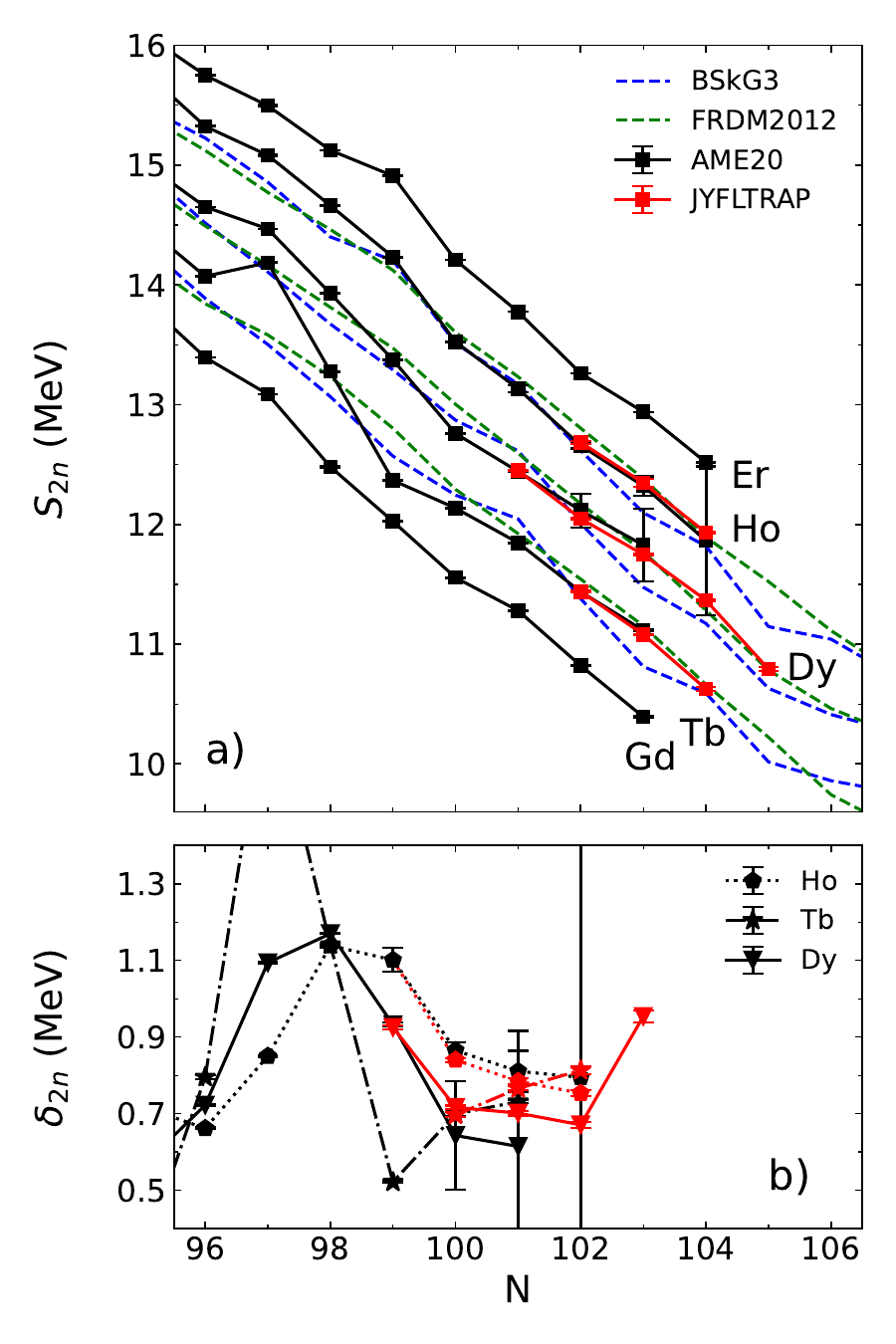}
\caption{a) Two-neutron separation energies $S_{2n}$ determined from the AME20 experimental mass values (in black), added together with the JYFLTRAP mass measurements from this work (in red) and the predictions from the BSkG3 mass model \cite{grams2023skyrme} (in dashed blue) and FRDM2012 \cite{moller2016nuclear} (in dashed green). b)  Differential $\delta_{2n}$ of the two-neutron separation energies for the isotopic chains studied in this work, identified with their own marker. AME20 experimental mass values are shown in black and the new and updated $\delta_{2n}$ values, determined from this work, are shown in red.\label{fig:s2n}}
\end{figure}

Two-neutron separation energies $S_{2n}$, defined as,
 \begin{equation}
     S_{2n}(Z,N)=\text{ME}(Z,N-2)-\text{ME}(Z,N)+2\text{ME}_n,
 \end{equation}
 with $\text{ME}(Z,N)$ being the mass excess of a nuclide with proton and neutron numbers $(Z,N)$, and  $\text{ME}_n$ the mass excess of a neutron, provide a way to probe changes in the nuclear structure. The isotopic chains ranging from praseodymium ($Z=59$) to dysprosium ($Z=66$) display a kink in the two-neutron separation energies at around $N=90-92$ explained by the onset of strong prolate deformation  \cite{brix1952nuclear,mottelson1955classification}. The deformation is predicted to reach a maximum around the midshell at $N\approx104$ \cite{grams2023skyrme,moller2016nuclear}, therefore it is interesting to explore the behavior of the $S_{2n}$ values in this region. Figure~\ref{fig:s2n}a shows the two-neutron separation energies $S_{2n}$ based on the mass-excess values from this work and the experimental values from AME20 \cite{AME2020}. In addition, the slope of the $S_{2n}$ values for the studied isotopic chains, defined as
 \begin{equation}
     \delta_{2n}(Z,N)=S_{2n}(Z,N)-S_{2n}(Z,N+2),
 \end{equation}
 is also presented, see Figure~\ref{fig:s2n}b. 
 
 With the high-precision mass values from this work we can pin down the trend more precisely than in the literature. For the studied isotopic chains, the $S_{2n}$ trends are rather smooth within their respective error bars and the mass values are compatible with a constant linear slope for $N=102-104$, see $\delta_{2n}(N=100-102)$ on Fig.~\ref{fig:s2n}b, for both the AME20 values \cite{AME2020} and the new mass measurements from this work. However, the $S_{2n}$ trend in the dysprosium isotopic chain gets steeper again crossing the midshell at $N=104$, see $\delta_{2n}(N=103)$, similarly to the Zr ($Z=40$) isotopic chain at the neutron midshell $N=66$ \cite{Hukkanen2024}. Moreover, the $\delta_{2n}$ values reach a maximum at $N=98$, which coincides with the minimum in the $2_1^+$ excitation energies (see Fig.~5 from \cite{Hartley2018}). The latter has been interpreted as due to a deformed subshell closure and maximum deformation reached at $N=98$ \cite{Hartley2018}. Another minimum in the $2_1^+$ energies is observed at $N=104$, indicating maximum deformation at the midshell that was also been predicted theoretically \cite{Regan2002}. Our new $\delta_{2n}$ values (Fig.~\ref{fig:s2n}b) reach a minimum at $N\approx100-102$ but increase steeply for $N>102$, suggesting that a similar local maximum as for $N=98$ can be reached at $N=104$. 
 
We have compared the experimental two-neutron separation energies $S_{2n}$ to two mass models in Fig.~\ref{fig:s2n}a. The Finite-Range Driplet Model 2012 (FRDM2012) \cite{moller2016nuclear} is commonly used in nuclear astrophysics calculations (see Sect.~\ref{sec:astro}) and has a rather good root-mean-square (rms) deviation of 0.3549~MeV for nuclei with $N\geq65$. The Brussels-Skyrme-on-a-Grid atomic mass model (BSkG3) \cite{grams2023skyrme}  
is a very recent model based on energy density functionals with a rms deviation of 0.631~MeV. Both models predict the studied nuclei to be strongly deformed with $\beta_2\approx 0.30$ \cite{moller2016nuclear} and $\beta_2\approx 0.35$ \cite{grams2023skyrme}, reaching the maximum in the $N=104$ region. While FRDM2012 follows a rather linear trend, the BSkG3 values have changes in the slope at certain neutron numbers. Interestingly, BSkG3 predicts a stronger decrease in the $S_{2n}$ values after $N=104$, followed by a rather flat trend from $N=105$ onwards for the studied isotopic chains. 

Despite the variations in the predicted values, the models agree with the experimental $S_{2n}$ values well, as can be seen from Fig.~\ref{fig:s2n}. A local average (ME$_\text{theory}-$ME$_\text{experiment}$) and rms deviation have also been calculated, considering 22 neutron-rich nuclei with $N\geq98$ in the studied isotopic chains. The local average for both FRDM and BSkG3 are positive, $0.048$~MeV and $0.423$~MeV, respectively, indicating on average the production of larger mass excesses compared to experimental values. On the other hand, the calculated rms deviations, $0.131$~MeV and $0.487$~MeV, respectively, are smaller than their overall rms deviations, showing the good performance of both models in the region. FRDM thus appears to agree better than BSkG3 with the selected data. Extending the mass measurements to $N=105$ also for the Tb and Ho isotopic chains would be important to test the theoretical predictions to further validate the models, and to probe if the observed slight slope change in the $S_{2n}$ dysprosium chain is a more general feature when crossing the neutron midshell at $N=104$.

\subsection{Impact on the astrophysical rapid neutron capture process}
\label{sec:astro}
The rare-earth abundance peak at $A\approx165$ is of special interest for the r-process studies. The solar-system r-process abundances in the rare-earth region are well known \cite{Cowan2021,goriely1999} and the region is produced via the main r process, which constrains the variety of astrophysical conditions that can produce these elements. 

Nuclear masses are one of the key inputs for the r process and the r-process abundances have been shown to sensitively depend on masses \cite{mumpower2016impact,Nikas2022}. The impact of the masses can be most directly seen in the neutron-capture reaction rates that serve as inputs for the r-process calculations. We have studied the impact of the mass values determined in this work on the astrophysical $(n,\gamma)$ neutron-capture reaction rates calculated with TALYS 1.96 \cite{talys}. The impact of the more precise mass values was further probed by varying the $Q$ value of the $(n,\gamma)$ using the min-max principle. 

\begin{figure}[h!t!b]
\includegraphics[width = 0.5\textwidth]{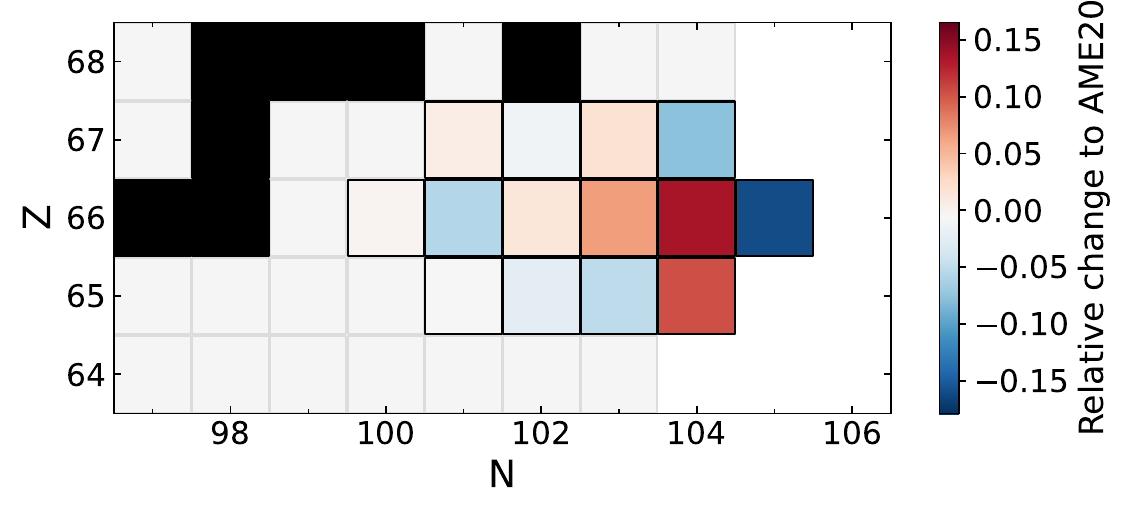}
\caption{The relative change of the astrophysical $^{A}_{N}X(n,\gamma)^{A}_{N+1}X$ reaction rate $R_\text{rel.}$ at $T=1$~GK, defined as $(R_\text{JYFL}-R_\text{AME20})/R_\text{AME20}$, where the rates were calculated using TALYS 1.96 with the mass values from this work ($R_\text{JYFL}$) and with the AME20  \cite{AME2020} values ($R_\text{AME20}$). The relative change magnitude of $R$ is color-coded in the tiles where reaction rates where calculated. The stable nuclides are shown as black tiles. \label{fig:rel_change}}
\end{figure}

Figure~\ref{fig:rel_change} shows the relative change between the reaction rates $R_\text{rel.}$ using the mass values from this work and the AME20 values \cite{AME2020}. They are calculated at a typical r-process temperature $T=1$~GK and defined as $(R_\text{JYFL}-R_\text{AME20})/R_\text{AME20}$.
The relative change on the reaction rates is most significant in the reactions involving masses measured for the first time in this work, as bigger discrepancies can be observed with respect to the estimations based on systematics. Hence, at $T=1$~GK, the new $^{169}$Tb$(n,\gamma)^{170}$Tb, $^{170}$Dy$(n,\gamma)^{171}$Dy, and $^{171}$Dy$(n,\gamma)^{172}$Dy, reaction rates are 9$\%$ higher, 16$\%$ higher, and 18$\%$ lower, respectively. 

The reaction rates obtained with the masses from this work and the AME20 experimental values all agree with each other, except for the $^{167}$Tb$(n,\gamma)^{168}$Tb reaction. While the discrepancy between the $^{168}$Tb mass from this work and the previous measurement \cite{vilen2020exploring} is relatively small, 34~keV, it is almost 6$\sigma$ away in terms of standard deviation considering the high precision given by Penning trap mass spectrometry. As a result, the new $^{167}$Tb$(n,\gamma)^{168}$Tb reaction rate at $T=1$~GK, only 2$\%$ lower, is yet about $3\sigma$ away from the previous value. It should be noted however that this effect remains negligible considering the other nuclear physics uncertainties at play in the reaction rates.

\begin{figure}[h!t!b]
\includegraphics[width = 0.5\textwidth]{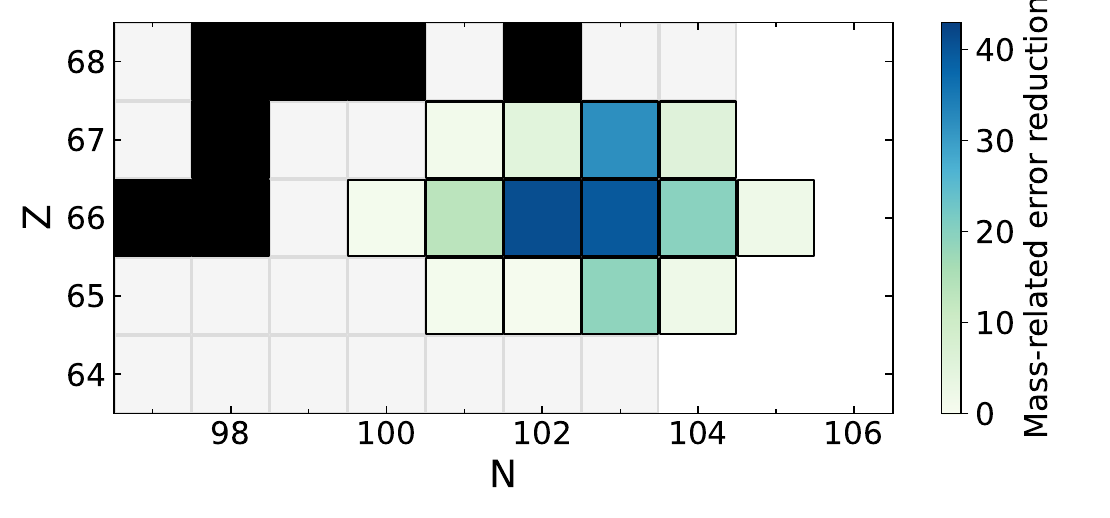}
\caption{The mass-related error reduction on the calculated astrophysical $^{A}_{N}X(n,\gamma)^{A}_{N+1}X$ reaction rates at $T=1$~GK, defined as $\delta_m R_\text{JYFL}/\delta_m R_\text{AME20}$, where the rates were calculated using TALYS 1.96 with the mass values from this work ($R_\text{JYFL}$) and with the AME20 \cite{AME2020} values ($R_\text{AME20}$). The reduction factor on $\delta_m$ is color-coded in the tiles where reaction rates where calculated. The stable nuclides are shown as black tiles.\label{fig:err_reduc} }
\end{figure}

The effect of the improved precision is shown in Fig.~\ref{fig:err_reduc}, where the reduction on the mass-related uncertainties in the calculated (n,$\gamma$) reaction rates $\delta_m R$, defined as $\delta_m R_\text{JYFL}/\delta_m R_\text{AME20}$, 
has been plotted. The biggest reduction of mass-related errors is seen in less exotic neutron captures, where both nuclear masses $A$ and $A+1$ of interest are experimentally known from mass spectrometry but with a limited precision. This leads, for instance, to the $^{168}$Dy$(n,\gamma)^{169}$Dy, $^{169}$Dy$(n,\gamma)^{170}$Dy, and $^{170}$Ho$(n,\gamma)^{171}$Ho reaction rates having their respective mass-related uncertainty decreased by a factor 43, 41, and 33, with our new measurements.

The newly determined reaction rates were implemented in the r-process nucleosynthesis calculations to study their impact on the r-process abundances. The r-process calculations performed started at a temperature of $T=10$~GK, where the medium is first characterized by a nuclear statistical equilibrium. The impact of the masses measured in this work on the final solar system r-process abundances in the rare-earth region were studied considering the same low-entropy r-process trajectory with $S=15$~$k_B$/baryon. The electron fraction was varied between $Y_e=0.12-0.2$ giving moderately neutron-rich ejecta, and the dynamic timescale was parameterized to $\tau=0.7$~ms, after which a homologous expansion was assumed. These astrophysical conditions are proposed to be met in neutron-star merger scenarios, see e.g. Refs.~\cite{RevModPhys.29.547,lippuner2015r,drout2017light}.

\begin{figure}[h!t!b]
\includegraphics[width = 0.45\textwidth]{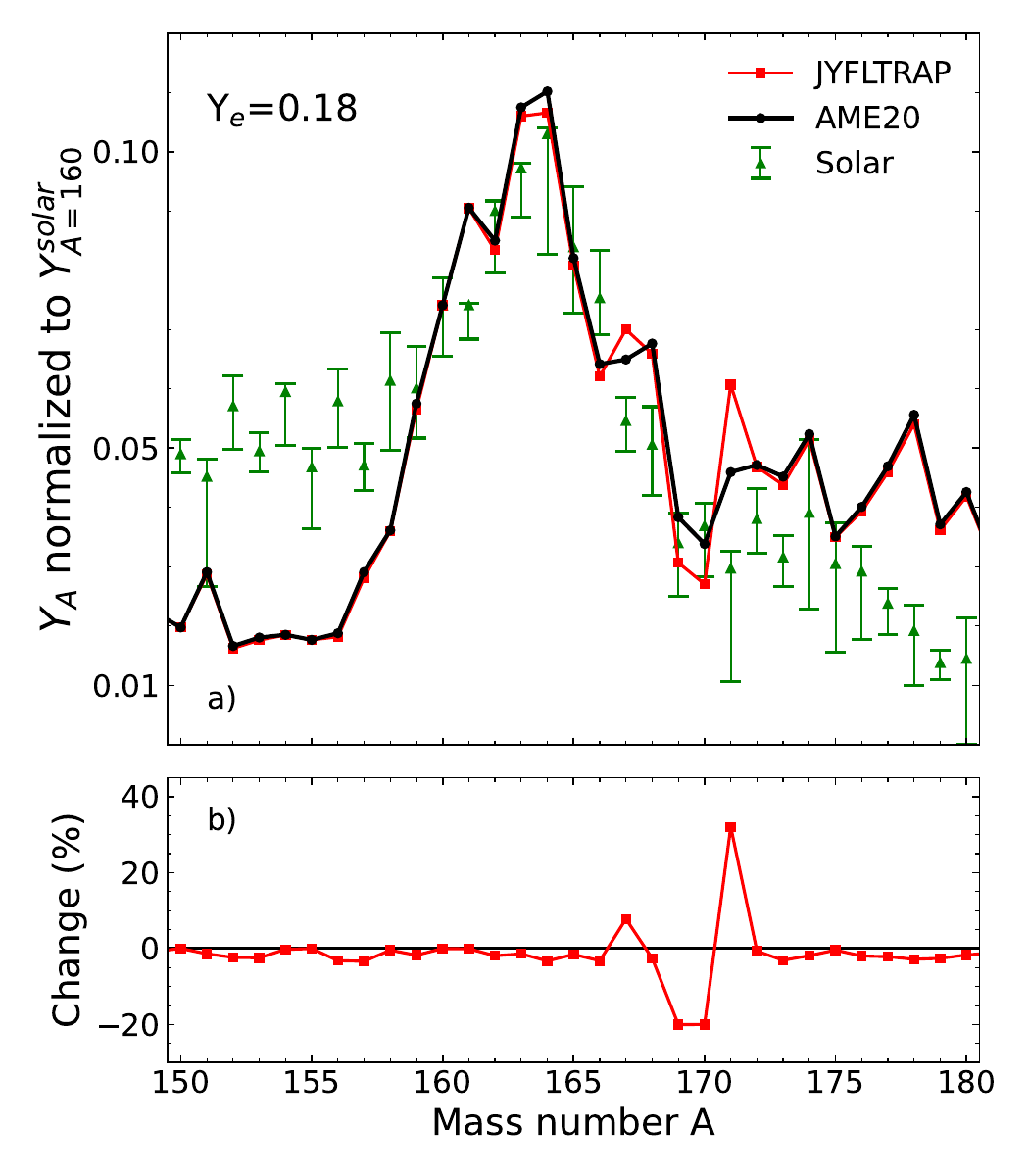}
\caption{a) r-process abundances $Y_A$ as a function of mass number A calculated using the mass values from this work (red) and from AME20 \cite{AME2020} (black), compared to the solar-system r-process abundances from Goriely et al. \cite{goriely1999} (green). The calculations were normalized with respect to the solar system abundance at A=160. b) Relative change in the calculated normalized abundances in $\%$, defined as $(Y_\text{norm,JYFL}-Y_\text{norm,AME20})/Y_\text{norm,AME20}$. The used trajectory considered an initial electron fraction $Y_e=0.18$ and an entropy per baryon of $S=15k_B$.
\label{fig:mass_plot} }
\end{figure}

Figure~\ref{fig:mass_plot} shows the final r-process abundance pattern obtained from the calculations using the results from this work, in comparison with a baseline calculation performed using the AME20 \cite{AME2020} mass values. For nuclei for which no AME20 experimental data existed, the extrapolated mass values based on systematics 
were used. The calculations were normalized to the closest r-only nuclide to the studied mass region, $^{160}$Gd, and compared with the solar system r-process abundances from Goriely et al. \cite{goriely1999}. 

In these astrophysical conditions, both sets of calculations reproduce the main part of the rare-earth abundance peak in the $A=160-165$ mass region rather well and the best fit is obtained with an initial $Y_e=0.18$. On the right side of the peak at higher mass numbers, however, discrepancies can be observed. Most significantly, our new values indicate a stronger minimum at $A=170$ and maximum at $A=171$ than the ones obtained with the baseline calculations using the AME20 values and extrapolations. The abundances are around $20\%$ lower at $A=169,170$ and $30\%$ higher at A=171 with new mass values from this work, reflecting the increased $(n,\gamma)$ rates at $A=169,170$ and decreased rates at $A=171$ (see Fig.~\ref{fig:rel_change}). These are mainly due to the changes in the mass-excess values of $^{171}$Dy ($-164$~keV) and $^{171}$Ho ($-90$~keV) with respect to the AME20 values. Similarly, with the $(n,\gamma)$ reaction rates at $A=167$ being decreased with our updated calculations, a higher abundance at $A=167$ is observed, with a change of around 8$\%$.

The r-process abundances around $A\approx170$ region calculated with the mass values from this work deviate more from the solar-system r-process residuals than the baseline calculations, see Fig.~\ref{fig:mass_plot}. As variations in nuclear masses affect all the relevant nuclear properties of neighboring nuclei, these variations might indicate a lack of reliable nuclear data in this region from the literature, or an overestimation of the observed solar system s-process abundances in the region that would affect the residual r-process abundances determination. 

\section{Summary and conclusions}
 We have measured atomic masses of 11 neutron-rich rare-earth isotopes with the JYFLTRAP double Penning trap at the IGISOL facility in the JYFL Accelerator Laboratory. Of these, $^{169}$Tb and  $^{170\text{,}171}$Dy mass values were experimentally determined for the first time. The neutron-rich $^{169\text{-}171}$Ho, as well as $^{169}$Dy, were explored for the first time via Penning-trap mass spectrometry. The precision of these mass values was greatly improved as compared to the literature values. The $^{167,168}$Tb and $^{167,168}$Dy mass values were in good agreement with existing Penning-trap data. For $^{168}$Tb, a significant discrepancy, attributed to the presence of an isobaric contaminant in the previously reported measurement, was identified.
 
The impact of our new mass measurements in the $N=104$ region on nuclear structure was studied via the two-neutron separation energy $S_{2n}$ and $\delta_{2n}$ parameters. The neutron midshell at $N=104$ was reached with $^{169}$Tb, $^{170}$Dy, $^{171}$Ho and crossed with $^{171}$Dy. Although a rather smooth behavior can be observed for the $S_{2n}$ values extended by our measurements, a subtle change in the $S_{2n}$ slope crossing the midshell at $N=104$ is noticed in the dysprosium chain. Further measurements would be needed to confirm if such a decrease after $N=104$ is a more general feature in the two-neutron separation energies. The experimental $S_{2n}$ values were in agreement with the predictions of the FRDM2012 \cite{moller2016nuclear} and BSkG3 \cite{grams2023skyrme} mass models.

The astrophysical implications of our results were also investigated, looking at the $(n,\gamma)$ reaction rates and the modelling of the rare-earth r-process peak in the solar abundance pattern. Our work shows the importance of using high precision mass spectrometry to improve the mass-related constraints given to the calculations. Moreover, pushing the mass measurements towards more exotic nuclei induced significant changes in the reaction rates that were previously partially based on theory and extrapolated values in the literature.

Calculations of r-process scenarios were performed to study the implications of our mass measurements on the r-process abundance pattern in the rare-earth region. Although the considered astrophysical trajectories give a relatively accurate modelling of the central region of the rare-earth peak, our new nuclear data display discrepancies with both the calculations based on the literature and the observed solar abundances at around mass $A=170$. These deviations suggest that the region needs to be further investigated by high-precision mass spectrometry but also by additional experimental methods able to access other uncertainties in neutron capture rates such as neutron-capture cross sections, and nuclear physics quantities like level structures, $\beta$-decay half-lives, $\beta$-delayed neutron-emission probabilities, relevant for the r-process. Such future studies will consolidate the input of nuclear data for theoretical mass models.

\begin{acknowledgments}
We acknowledge the funding from the European Union’s Horizon 2020 research and innovation programme under Grant Agreement No. 771036 (ERC CoG MAIDEN) and No. 861198–LISA–H2020-MSCA-ITN-2019, from the European Union’s Horizon Europe Research and Innovation Programme under Grant Agreement No. 101057511 (EURO-LABS) and from the Research Council of Finland projects No. 354589, No. 295207, No. 306980, No. 327629 and No. 354968. The financial support provided by the Vilho, Yrjö and Kalle Väisälä Foundation is acknowledged by J. Ruotsalainen. 

\end{acknowledgments}

\bibliography{bibliography}

\end{document}